\documentclass{PoS}

\usepackage{tabularx}

\usepackage[utf8]{inputenc}
\usepackage[T1]{fontenc}
\usepackage[normalem]{ulem}
\usepackage[english]{babel}

\usepackage{tabularx}
\usepackage{array}
\usepackage{graphics}
\usepackage[pdftex]{graphicx}
\usepackage{epsfig}
\usepackage{amsmath}
\usepackage{amssymb}
\usepackage{bbold}

\usepackage{soul}
\usepackage{graphics}

\usepackage{cite}

\usepackage{ae}
\usepackage{mathtools}
\usepackage{amsfonts}

\usepackage{pifont}% http://ctan.org/pkg/pifont
\newcommand{\xmark}{\ding{55}}%

\usepackage{enumerate}

\usepackage{bbm}
\usepackage{bm}
\usepackage{makecell}

\usepackage{arydshln}

\title{Theory overview of semileptonic $B$ decays}

\ShortTitle{Theory overview of semileptonic $B$ decays}

\author{\speaker{Olcyr Sumensari}%\thanks{A footnote may follow.}

{\par\centering \vskip 0.1 cm\par}
{\sl 
Istituto Nazionale Fisica Nucleare, Sezione di Padova, I-35131 Padova, Italy}\\
{\par\centering \vskip -0.3 cm\par}
{\sl 
Dipartamento di Fisica e Astronomia ``G.~Galilei", Università di Padova, Italy}\\
        E-mail: \email{olcyr.sumensari@pd.infn.it}}

\abstract{I review the recent theoretical progress on the study of tree-level semileptonic $b\to c$ decays. After briefly reviewing the latest developments on the exclusive determinations of $|V_{cb}|$, I discuss the hints on lepton flavor universality violation in $B\to D^{(\ast)}\ell \bar{\nu}$ decays. Particular emphasis is given to the implications of these anomalies and to the possibilities to experimentally test the proposed New Physics explanations. }

\FullConference{The International Conference on B-Physics at Frontier Machines - BEAUTY2018\\
		6-11 May, 2018\\
		La Biodola, Elba Island, Italy}

\begin{document}

\section{Introduction}

Tree-level semileptonic decays of mesons provide a straightforward way to extract the moduli of several CKM matrix elements, which can then be used to test the CKM matrix unitarity. In the last decade, precision studies of semileptonic $B$-meson decays have been made possible by the the large samples of $B$-mesons collected at the $B$-factories (BaBar and Belle) and at LHCb. At the same time, we witnessed a considerable progress in numerical simulations of QCD on the lattice (LQCD), which nowadays allow us to attain a percent level precision for certain hadronic quantities. For these reasons, semileptonic $B$ decays offer a very promising route to search for the effects of New Physics (NP). 

One of the most intriguing results from the $B$-factories and LHC is the indication of lepton flavor universality (LFU) violation, observed in both tree-level and loop-induced semileptonic $B$-decays. The $B$-physics experiments observed a departure from the SM in the tree-level processes, i.e.~those mediated by the charged currents, namely,

%%%%%%%%%%%%%%%%
\begin{equation}
\label{eq:RD-RDst}
R_{D^{(\ast)}} = \dfrac{\mathcal{B}(B\to D^{(\ast)}\tau \bar{\nu})}{\mathcal{B}(B\to D^{(\ast)}l \bar{\nu})}\,,\qquad\qquad l \in \lbrace e,\mu\rbrace\,,
\end{equation}
%%%%%%%%%%%%%%%% 

\noindent which turned out to be $\approx 2-3\,\sigma$ \underline{above} the SM predictions~\cite{Lees:2012xj,Lees:2013uzd,Huschle:2015rga,Aaij:2015yra,Hirose:2016wfn,Sato:2016svk,Abdesselam:2016cgx}. When combined in the same fit, these results amount to a discrepancy with respect to the SM at the $\approx 4\,\sigma$ level~\cite{Amhis:2016xyh}. This observation is corroborated by the first measurement of the ratio $R_{J/\Psi}=\mathcal{B}(B_c \to J/\Psi \tau \bar{\nu})/\mathcal{B}(B_c \to J/\Psi l \bar{\nu})$~\cite{Aaij:2017tyk}, which also appears to be larger than the SM estimates~\cite{RJpsipaper}. Another intriguing indication of LFU violation appeared in loop-induced processes, 

%%%%%%%%%%%%%%%%
\begin{equation}
\label{eq:RK-RKst}
R_{K^{(\ast)}} = \dfrac{\mathcal{B}(B\to K^{(\ast)} \mu\mu)}{\mathcal{B}(B\to K^{(\ast)}ee)}\Bigg{|}_{q^2\in [q^2_{\mathrm{min}},q^2_{\mathrm{max}}]}\,,
\end{equation}
%%%%%%%%%%%%%%%% 

\noindent measured by LHCb in different $q^2$-bins below the $c\bar{c}$ resonances~\cite{Aaij:2014ora,Aaij:2017vbb}. The obtained values appear to be $\approx 2.5\,\sigma$ \underline{below} the SM predictions \cite{Hiller:2003js,Bordone:2016gaq}, amounting to a combined discrepancy at the $\approx 4\sigma$ level. 

Both types of ratios are defined in such a way that the CKM matrix element dependence and most of the hadronic uncertainties cancel-out, which makes then rather clean observables. Therefore, if confirmed with more data, they will represent a very clear indication of NP. 

\

In the following we will focus on two issues that recently attracted the attention of the theory community: (i) the extraction of $|V_{cb}|$ from $B\to D^\ast l \bar{\nu}$ decays, and (ii) the above-mentioned hints of LFU violation in charged-current processes.~\footnote{The reader is referred to other talks of the same conference for a discussion of the LFU anomalies in neutral currents~\cite{othertalks}.}

\section{Exclusive determinations of $|V_{cb}|$: latest developments}
\label{sec:Vcb}

The CKM matrix element $|V_{cb}|$ is a free parameter of the SM that plays an important role in the unitarity triangle analysis and in the prediction of flavor changing neutral currents~\cite{Charles:2015gya,Bona:2006ah}. Its value is extracted by confronting the experimental results for the processes based on the transition $b\to c l\bar{\nu}$ with their SM predictions. Surprisingly, the values determined by using inclusive and exclusive $B$-meson decays show a discrepancy that only grew with time. The current averages quoted by HFLAV are~\cite{Amhis:2016xyh}
%%%%%%%%%%%%%
\begin{align}
 |V_{cb}|^{\mathrm{incl.}} &=(42.19\pm 0.78)\times 10^{-3}\,,\qquad\qquad\quad\qquad\qquad \text{from} \quad B\to X_c l \bar{\nu}\,,\\[0.3em]%
|V_{cb}|^{\mathrm{excl.}} &=(39.05\pm 0.47_{\mathrm{exp}}\pm 0.58_{\mathrm{th}})\times 10^{-3}\,, \qquad\quad\quad\hspace*{0.38em} \text{from} \quad B\to D^\ast l \bar{\nu}\,,\\[0.3em]
|V_{cb}|^{\mathrm{excl.}}  &=(39.18\pm 0.94_{\mathrm{exp}}\pm 0.36_{\mathrm{th}})\times 10^{-3}\,, \qquad\quad\quad\hspace*{0.38em} \text{from} \quad B\to D l \bar{\nu}\,,
\end{align}
%%%%%%%%%%%%%

\noindent which differ by $\approx 3 \sigma$. To reliably extract these values it is crucial to control the hadronic uncertainties.~\footnote{The relevant $B\to D^{(\ast)}$ hadronic matrix elements are expressed in terms of two (four) form factors in the SM: $\langle D (k) \vert \bar{c} \gamma^\mu P_L b \vert B(p)\rangle \propto f_+(q^2),f_0(q^2)$ and  $\langle D^\ast (k) \vert \bar{c} \gamma^\mu P_L b \vert B(p)\rangle \propto V(q^2),A_1(q^2),A_2(q^2),A_0(q^2)$, where $q^2=(p-k)^2$ is the dilepton squared mass. Moreover, the scalar and pseudoscalar form factors, $f_0(q^2)$ and $A_0(q^2)$, only contribute significantly to $B\to D^{(\ast)}\tau \bar{\nu}$.} While the relevant $B\to D$ form factors have been computed at nonzero recoil values by means of LQCD simulations~\cite{Na:2015kha,Lattice:2015rga}, cf.~discussion in Sec.~\ref{sec:RD-sm}, some pieces of information on the $B\to D^\ast$ form factors are still lacking. The usual strategy in the latter case is to employ the overall form-factor normalization $[h_{A_1}(1)]$ computed on the lattice~\cite{Bailey:2014tva,Harrison:2016gup} and then extract their shapes from the experimental angular distributions of $B\to D^\ast(\to D\pi)l\bar{\nu}$, by using a convenient parameterization. The parameterization considered by the experimental collaborations is the one proposed by Caprini, Lellouch and Neubert (CLN)~\cite{Caprini:1997mu}, which suggests
%%%%%%%%%%%%
\begin{align}
\label{eq:hA1}
h_{A_1}(w) &= {h_{A_1}(1)}\left[1+8 {\rho^2} z +(53 {\rho^2}-15)z^2-(231 {\rho^2}-91)z^3\right]\,,\\[0.3em]
\label{eq:R1}
R_1(w) &= {R_1(1)}{-0.12} (w-1){+0.05}(w-1)^2\,,\\[0.3em]
\label{eq:R2}
R_2(w) &= {R_2(1)}{-0.11} (w-1){-0.06}(w-1)^2\,,
\end{align}
%%%%%%%%%%%%

\noindent where $w=(m_B^2+m_D^2-q^2)/(2 m_B m_D)$ is the relative velocities of the initial and final state mesons, and $z=(\sqrt{w+1}-\sqrt{2})/(\sqrt{w+1}+\sqrt{2})$. This parameterization is based on two approximations: 
\begin{enumerate}
\item[(i)] The shape of the form factor $h_{A_1}(w)$, which corresponds to the Isgur-Wise function in the heavy-quark limit, is truncanted after the quadratic term in $w-1$. Unitarity and analyticity are then used to express the shape parameters in Eq.~\eqref{eq:hA1} in terms of a single parameter $\rho$, the slope of the Isgur-Wise function, cf.~Ref.~\cite{Caprini:1995wq,Caprini:1997mu,Boyd:1997kz}; 
\item[(ii)] The ratios $R_{1,2}(w)$ between other form factors and $h_{A_1}(w)$ are also parameterized by a power expansion around zero-recoil values ($w=1$), with their respective shapes and curvatures fixed to phenomenologically plausible values with no error bars.~\footnote{It is worth to stress that the accuracy of the approximation in Eq.~\eqref{eq:R1} and \eqref{eq:R2} was estimated to be better than $2\%$ in Ref.~\cite{Caprini:1995wq}. The uncertainty associated to the shape parameters, negligible at the time, are now important and need to be included in experimental analyses, cf.~discussion below on the BGL parameterization.} 
\end{enumerate}
The overall normalization $[h_{A_1}(1)]$ is then determined on the lattice~\cite{Bailey:2014tva,Harrison:2016gup}, while the three remaining parameters in \eqref{eq:hA1}--\eqref{eq:R2}, namely $\rho$ and $R_{1,2}(1)$, are extracted from the $B\to D^\ast l \bar{\nu}$ data, cf.~e.g.~Ref.\cite{Amhis:2016xyh}.

Recently, the published unfolded $B\to D^\ast l \bar{\nu}$ distributions by Belle~\cite{Abdesselam:2017kjf} allowed theorists to perform their own extraction of $|V_{cb}|^{\mathrm{excl.}}$ by using slightly different parameterizations. The authors of Refs.~\cite{Bigi:2017njr,Grinstein:2017nlq} independently concluded that the obtained value of $|V_{cb}|^{\mathrm{excl.}}$ is parameterization dependent. The main difference is that these new authors considered the slopes and curvatures of $R_{1,2}(w)$ free parameters to be fixed from the fit with data, following a proposal by Boyd, Grinstein and Lebed (BGL)~\cite{Boyd:1997kz}. The results obtained with both parameterizations read~\cite{Bigi:2017njr,Grinstein:2017nlq} 
%%%%%%%%%%%%
\begin{align}
|V_{cb}|^{\mathrm{excl.}}_{\mathrm{CLN}} =  (38.2\pm 1.5) \times 10^{-3}\,,\qquad\qquad |V_{cb}|^{\mathrm{excl.}}_{\mathrm{BGL}} = 41.7^{+2.0}_{-2.1} \times 10^{-3}\,.
\end{align}
%%%%%%%%%%%%

\noindent Interestingly, the values obtained with BGL are closer to the inclusive value.~\footnote{Consistent results have also been very recently reported by the Belle collaboration \cite{belle-ichep}} While the BGL parameterization seems to be more attractive, since it does not rely on HQET relations, we still cannot conclude that one parameterization is better than another. Both fits provide an equally good description of current data. For that reason, we still need to wait for LQCD and Belle-II data at small-recoil values to finally solve the $V_{cb}$ problem.

\section{Lepton flavor universality violation: $R_{D}$ and $R_{D^{\ast}}$}

\subsection{Current status}
\label{sec:RD-sm}

We shall now discuss the hints of lepton flavor universality in tree-level (charged-current) $B$-decays, which are independent of the above discussion since $h_{A_1}(1) |V_{cb}|$ cancel out in these observables. The experimental averages of the LFUV ratios $R_{D^{(\ast)}}$ are given by~\cite{Amhis:2016xyh}
%%%%%%%%%%%%
\begin{align}
R_D^{\mathrm{exp}}= \dfrac{\mathcal{B}(B\to D\tau \bar{\nu})}{\mathcal{B}(B\to D l \bar{\nu})}=0.41(5)\,,\qquad\quad R_{D^\ast}^{\mathrm{exp}}= \dfrac{\mathcal{B}(B\to D^{\ast}\tau \bar{\nu})}{\mathcal{B}(B\to D^{\ast}l \bar{\nu})}=0.304(15)\,,
\label{RD_RDstar_exp}
\end{align}
%%%%%%%%%%%%%

\noindent to be compared with the SM predictions $R_D^{\mathrm{SM}}=0.299(7)$~\cite{Na:2015kha,Lattice:2015rga} and $R_{D^\ast}^{\mathrm{SM}}=0.257(3)$~\cite{Bernlochner:2017jka}, which are respectively $\approx 2\,\sigma$ and $\approx 3\,\sigma$ larger than the experimental values given above.~\footnote{Other predictions found in the literature are $R_{D^\ast}^{\mathrm{SM}}=0.252(3)$~\cite{Fajfer:2012vx}, $R_{D^\ast}=0.257(5)$~\cite{Jaiswal:2017rve} and $R_{D^\ast}^{\mathrm{SM}}=0.260(8)$~\cite{Bigi:2017jbd}, in agreement with the values quoted above.} As anticipated in the introduction, the main advantage of using ratios as in the equations above is that many hadronic uncertainties actually cancel, providing theoretically clean observables. Nonetheless, there are still residual hadronic uncertainties that must be carefully estimated. An important ingredient in the computation of $R_{D^{(\ast)}}$ is that the $\tau$-lepton mass is not negligible in comparison to $m_B$. Therefore, the (pseudo)scalar form factors, which do not contribute to the rates with light leptons, are needed to reliably predict these quantities. In the case of $R_D$ both scalar, $f_0(q^2)$, and vector, $f_+(q^2)$, form factors  have been computed on the lattice in the region of large $q^2$ by two different collaborations~\cite{Na:2015kha,Lattice:2015rga}. Extrapolation to $q^2=0$ is highly constrained by the relation $f_0(0)=f_+(0)$. The situation is slightly less favorable in the case of $R_{D^\ast}$ since the needed form factors are not yet available from LQCD simulations at nonzero recoils. The strategy adopted is to extract the leading form factors from $B\to D^\ast l \bar{\nu}$, as explained in  Sec.~\ref{sec:Vcb}. The remaining pseudoscalar form factor, $A_0(q^2)$, which cannot be extracted from data is then obtained by considerations based on HQET with generous error bars~\cite{Bernlochner:2017jka,Bigi:2017jbd}. The overall agreement in the literature is that the large departures from the SM cannot be explained by only relying on underestimated hadronic uncertainties. This conclusion is mostly based on the fact that the (unknown) $A_0(q^2)$ form factor gives a numerically small contribution to $R_{D^\ast}$ but still not negligible. It goes without saying that its computation on the lattice would be very welcome. Finally, it is worth mentioning that another potential source of uncertainty comes from the soft-photon radiation, which have been partially estimated in Ref.~\cite{Becirevic:2009fy}.

\subsection{Effective field theory description}
\label{sec:eft}

In the following, we will assume that the $R_{D^{(\ast)}}$ anomalies are due to NP and we will discuss their implications. The transition $b\to c \ell \bar{\nu}$ (with $\ell=e,\mu,\tau$) can be generically described at low-energies by the following effective Lagrangian, at dimension-6, 
%%%%%%%%%%%%
\begin{align}
\label{leff:bc}
\mathcal{L}_{\mathrm{eff}} = -2 \sqrt{2}& G_F V_{cb} \bigg{[}(1+g_{V_L}^{\ell})\,\big{(}\bar{c}_L \gamma^\mu b_L\big{)}\big{(}\bar{\ell}_{L} \gamma_\mu \nu_{L}\big{)} +g_{V_R}^{\ell}\,\big{(}\bar{c}_R \gamma^\mu b_R\big{)}\big{(}\bar{\ell}_{L} \gamma_\mu \nu_{L}\big{)}\\
& + g_{S_R}^{\ell}\,\big{(}\bar{c}_L  b_R\big{)}\big{(}\bar{\ell}_{R} \nu_{L}\big{)}+ g_{S_L}^{\ell}\,\big{(}\bar{c}_R b_L\big{)}\big{(}\bar{\ell}_{R} \nu_{L}\big{)}+ g_T^{\ell}\, \big{(}\bar{c}_R \sigma^{\mu\nu} b_L\big{)}\big{(}\bar{\ell}_{R} \sigma_{\mu\nu} \nu_{L}\big{)}\bigg{]}+\mathrm{h.c.,}\nonumber
\end{align}
%%%%%%%%%%%%

\noindent where $g_{V_{L(R)}}^\ell$, $g_{S_{L(R)}}^\ell$ and $g_T^\ell$ are generic Wilson coefficients. Since the NP responsible for this interaction must arise above the electro-weak scale, one should further impose $SU(2)_L\times U(1)_Y$ gauge invariance to the above Lagrangian~\cite{Buchmuller:1985jz,Grzadkowski:2010es}. The operators one should then consider are~\cite{Aebischer:2015fzz}
%%%%%%%%%%%%%
\begin{align}
\label{eq:su2-operators}
\begin{split}
\big{[}\mathcal{O}_{\ell q}^{(3)}\big{]}_{prst} &=  \big{(}\overline{L_p}\,\gamma_\mu \tau^I \,L_{r}\big{)} \big{(}\overline{Q_{s}} \,\gamma^\mu \tau^I \,{Q_t}\big{)}\,,\quad\qquad
\big{[}\mathcal{O}_{\ell e qu}^{(1)}\big{]}_{prst} = \big{(}\overline{L^a_p}\, e_{rR}\big{)} \varepsilon_{ab} \big{(}\overline{Q^b_s} \,  u_{tR}\big{)}\,,\\[0.3em]
\big{[}\mathcal{O}_{\ell e dq}\big{]}_{prst} &=  \big{(}\overline{L^{a}_p} \,e_{rR}\big{)} \big{(}\overline{d_{sR}}  \,{Q_t}^a\big{)}\,,\quad\qquad\hspace*{2.9em}
\big{[}\mathcal{O}_{\ell e qu}^{(3)}\big{]}_{prst} = \big{(}\overline{L_p^a}\, \sigma_{\mu\nu}\,e_{rR}\big{)} \varepsilon_{ab} \big{(}\overline{Q^b_s}\, \sigma^{\mu\nu} \,u_{Rt}\big{)}\,,
\end{split}
\end{align} 
%%%%%%%%%%%%%

\noindent where $a,b$ are $SU(2)_L$ indices, $\varepsilon_{12}=-\varepsilon_{21}=1$, and $p,r,s,t$ are flavor indices. The matching at $\mu=m_b$ of this basis of operators onto the $SU(3)_c \times U(1)_\mathrm{em}$ invariant one, given in Eq.~\eqref{leff:bc}, tell us that $g_{V_R}^\ell$ cannot break LFU at dimension-6. This coefficient can only be generated by the LFU conserving operator $\mathcal{O}_{H u}= \big{(}H^\dagger i \overleftrightarrow{D}_\mu H\big{)}\big{(}\bar{u}_{pR} \gamma^\mu u_{rR}\big{)}$, being irrelevant for the following discussion~\cite{Fajfer:2012jt,Aebischer:2015fzz,Bernard:2006gy}. The other Wilson coefficients have a one-to-one correspondence to the basis given above.

To identify the allowed combination of Wilson coefficients at $\mu=m_b$, one should perform a fit to the experimental values in Eq.~\eqref{eq:RD-RDst}. Our working assumption will be that the NP couplings to light leptons are negligble, as suggested by current flavor data. The compact expressions for these observable are then given by~\cite{Feruglio:2018fxo}
%%%%%%%%%%%%%
\begin{align}
\label{eq:RD-RDst-compact}
\begin{split}
\dfrac{R_{D^{(\ast)}}}{R_{D^{(\ast)}}^{\mathrm{SM}}} =  |1+g_{V_L}^\tau|^2 &+ a^{D^{(\ast)}}_{S} \, |g_S^\tau|^2 + a^{D^{(\ast)}}_{P} \, |g_P^\tau|^2+ a_T^{D^{(\ast)}} \, |g_T^\tau|^2 + a^{D^{(\ast)}}_{SV_L}\,\mathrm{Re}\left[g_S^\tau \left(1+\left(g_{V_L}^{\tau}\right)^\ast \right)\right]\\[0.3em]
&+ a^{D^{(\ast)}}_{PV_L}\,\mathrm{Re}\left[g_P^\tau\left(1+\left(g_{V_L}^{\tau}\right)^\ast \right)\right]+ a_{TV_L}^{D^{(\ast)}}\,\mathrm{Re}\left[g_T^\tau\left(1+\left(g_{V_L}^{\tau}\right)^\ast \right)\right]\,,
\end{split}
\end{align}
%%%%%%%%%%%%

\noindent where $g_{S(P)}^\tau = g_{S_R}^\tau \pm g_{S_L}^\tau$, and the coefficients $a_i^D$ and $a_i^{D^\ast}$ are collected in Table~\ref{tab:RD-RDst-compact}. Interestingly, $R_D$ and $R_{D^\ast}$ are sensitive to a complementary set of NP operators. While $R_D$ is sensitive to both scalar and tensor contributions, $R_{D^\ast}$ only receives sizable contributions from the tensor ones. Furthermore, the scenario with only $g_{V_L}$ predicts $R_D/R_D^{\mathrm{SM}}=R_{D^\ast}/R_{D^\ast}^{\mathrm{SM}}$, since it corresponds to an overall shift of the SM. In this case, a viable explanation of $R_{D^{(\ast)}}$ can be obtained for
%%%%%%%%%%%%%
\begin{align}
\label{eq:gvl-fit}
g_{V_L}^\tau \in (0.09,0.13)\,.
\end{align}
%%%%%%%%%%%%%
to $1\sigma$ accuracy. Other viable solutions, which are not in the (current)$\times$(current) form, are not only possible but also well motivated by concrete NP scenarios, as we will discuss in Sec.~\ref{sec:concrete}. The possibility of considering (pseudo)scalar operators (i.e.~only $g_{S_{L(R)}}\neq 0$) was considered in the past, but it has been recently shown that such a scenario containing is in strong tension with the $B_c$-meson lifetime constraint~\cite{Alonso:2016oyd,Celis:2016azn}. More specifically, the pseudoscalar interactions lift the helicity suppression of the SM rate for

\begin{figure}[ht!]
\centering
\includegraphics[width=0.5\linewidth]{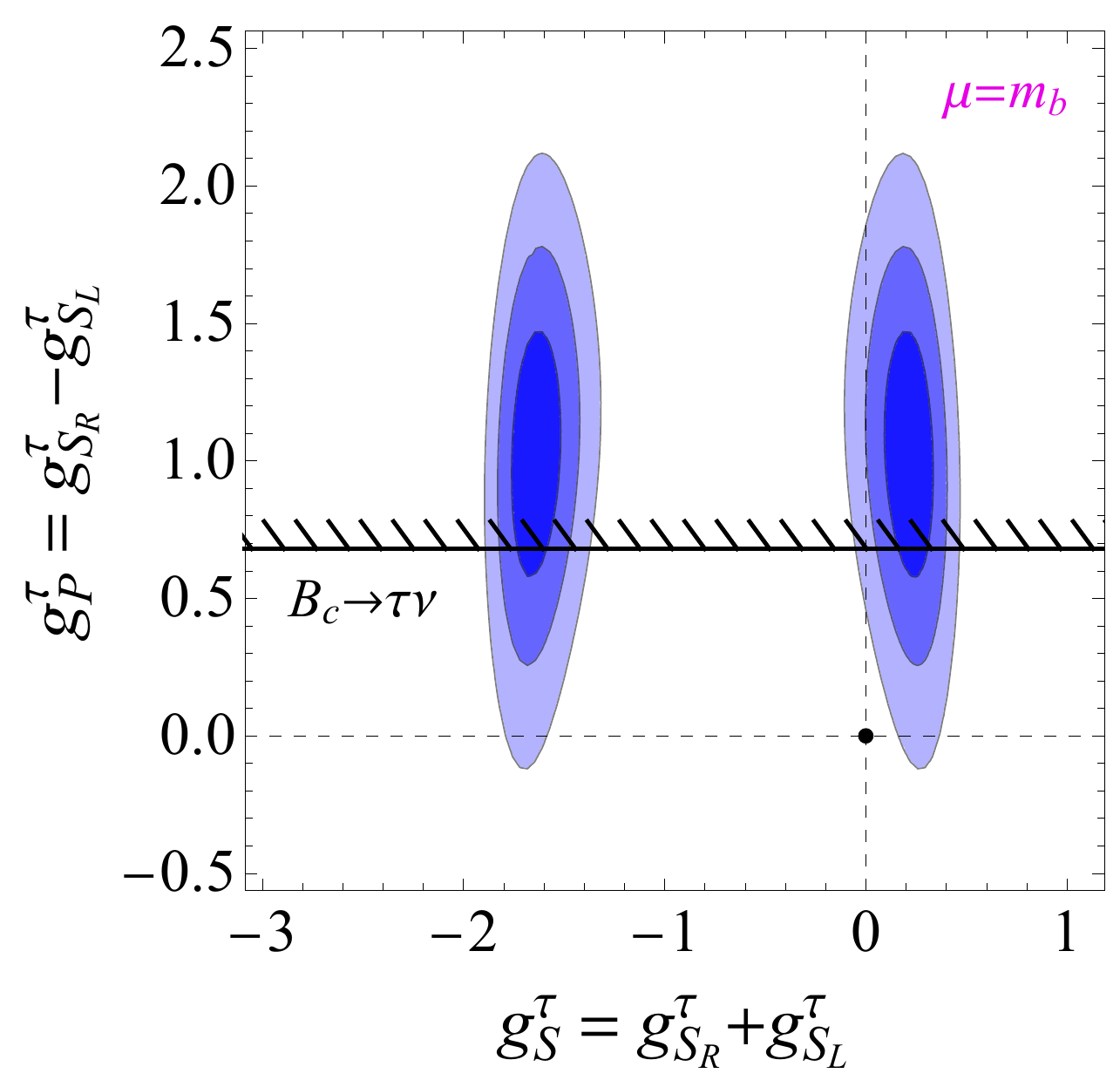}~\includegraphics[width=0.5\linewidth]{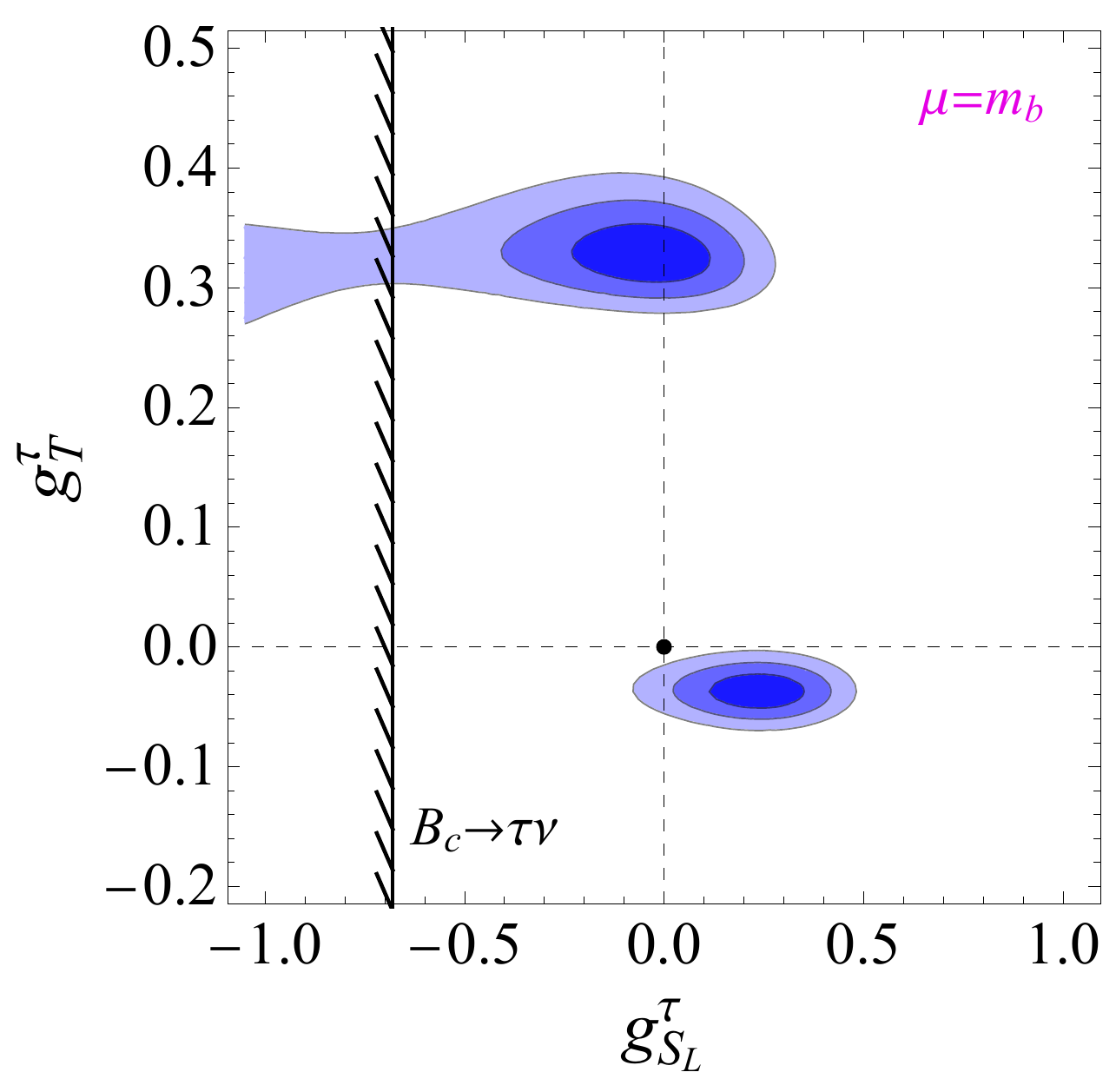}
\caption{\small \sl Allowed regions by $R_D$ and $R_{D^\ast}$ in the planes $g_{S}^\tau$ vs.~$g_P^\tau$ (left panel), 
and $g_{S_L}^\tau$ vs.~$g_T^\tau$ (right panel) are shown to $1$, $2$ and $3\,\sigma$ accuracy in blue (darker to lighter). The black lines show the constraints from the $B_c$--meson lifetime, which allow us to exclude the solutions with large values of $|g_P|$, as explained in the text \cite{Feruglio:2018fxo}.}
\label{fig:RD-RDst-1TeV}
\end{figure}

%%%%%%%%%%%%%%%%%%%%%
\begin{table}[t!]
\renewcommand{\arraystretch}{1.4}
\centering
\begin{tabular}{|c|cccccc|}\hline
Decay mode  & $a_S^M$ & $a_{SV_L}^M$ & $a_P^M$ & $a_{PV_L}^M$ & $a_T^M$ & $a_{TV_L}^M$\\ \hline\hline 
$B \to D$	&	 $1.08(1)$   &     $1.54(2)$      &  $0$	&	$0$ & $0.83(5)$	& $1.09(3)$\\
$B\to {D^\ast}$	&	  $0$  &    $0$   & $0.0473(5)$	&	$0.14(2)$ &	$17.3(16)$ & $-5.1(4)$ \\ \hline
\end{tabular}
\caption{ \sl \small Numerical values of the coefficients $a_{i}^M$ defined in Eq.~\eqref{eq:RD-RDst-compact} for $M\in \lbrace D,D^\ast \rbrace$ and $i \in \lbrace S, S V_L, P, P V_L, T, T V_L\rbrace$, assuming that NP only induces nonzero values of $g_{V_L}^\tau$, $g_{S_{L(R)}}^\tau$ and $g_T^\tau$, i.e.~the effective couplings to light leptons is negligible~\cite{Feruglio:2018fxo}. }
\label{tab:RD-RDst-compact} 
\end{table}
%%%%%%%%%%%%%%%%%%%%

%%%%%%%%%%%%%%%%%
\begin{equation}
\mathcal{B}(B_c\to \tau \bar{\nu}) = \tau_{B_c} \dfrac{m_{B_c} f_{B_c}^2 G_F^2 |V_{cb}|^2}{8 \pi} m_\tau^2 \left( 1- \dfrac{m_\tau^2}{m_{B_c}^2} \right)^2 \Bigg{|} 1 + g_{P}^\tau \dfrac{m_{B_c}^2}{m_\tau (m_b+m_c)}\Bigg{|}^2\,,
\end{equation}
%%%%%%%%%%%%%%%%%

\noindent where $f_{B_c}=427(6)$~MeV is the $B_c$--meson decay constant~\cite{McNeile:2012qf}. The current experimental value for $\tau_{B_c}= 0.507(9)$~ps~\cite{Patrignani:2016xqp} allows us to set a conservative limit of $30\%$ on $\mathcal{B}(B_c\to \tau \bar{\nu})$~\cite{Alonso:2016oyd}, which can then be translated onto the $1\,\sigma$ bound, $g_P^\tau (\mu=m_b) \in (-1.14,0.68)$, where we have used $|V_{cb}|=0.0417(20)$~\cite{Bigi:2017njr,Grinstein:2017nlq}.~\footnote{Alternatively, one could consider the less conservative limit $\mathcal{B}(B_c \to \tau \bar{\nu}) \lesssim 10\%$ as proposed in Ref.~\cite{Akeroyd:2017mhr}. By using this constraint instead, the bound would become $g_P^\tau\in (-0.76,0.30)$ to $1\,\sigma$ accuracy.} This constraint is illustrated in the left panel of Fig.~\ref{fig:RD-RDst-1TeV}, where it can be seen that an explanation of $R_{D^{(\ast)}}$ via only (pseudo)scalar operators is ruled out. Scenarios providing a good fit to $R_{D^{(\ast)}}$, while avoiding the $B_c$-lifetime constraint, can be obtained by using a different set of operators. For instance, a scenario containing only $g_{S_L}$ and $g_T$ can provide a good description of current data, as illustrated in the right panel of Fig.~\ref{fig:RD-RDst-1TeV}. This combination of Wilson coefficient is well motivated by scenarios containing leptoquark bosons which predict certain correlations between $g_{S_L}$ and $g_T$. Furthermore, it is known that a non-negligible mixing of $g_T$ into $g_{S_L}$ is induced via electroweak quantum corrections~\cite{Gonzalez-Alonso:2017iyc}, cf.~also Ref.~\cite{Feruglio:2018fxo}. Finally, note that viable scenarios with complex Wilson coefficients have also been considered in the literature, providing an equally good fit to the experimental results~Ref.~\cite{Sakaki:2013bfa,Hiller:2016kry,Becirevic:2018uab,Becirevic:2018afm}.

%%%%%%%%%%%%%%%%%%%%%%%%%%%%%%%%%%%%%%%%%
%%%%%%%%%%%%%%%%%%%%%%%%%%%%%%%%%%%%%%%%%
%%%%%%%%%%%%%%%%%%%%%%%%%%%%%%%%%%%%%%%%
%%%%%%%%%%%%%%%%%%%%%%%%%%%%%%%%%%%%%%%%

\subsection{Concrete New Physics scenarios}
\label{sec:concrete}

Several concrete models of NP can generate the viable Wilson coefficients discussed above. These scenarios require new bosonic degrees of freedom at the TeV scale~\cite{DiLuzio:2017chi} with tree-level contributions to the transition $b\to c\tau \bar{\nu}$. The main challenge is to comply with nontrivial constraints coming flavor physics observables, electroweak precision tests~\cite{Feruglio:2016gvd,Feruglio:2017rjo,Cornella:2018tfd} and the direct searches at the LHC~\cite{Faroughy:2016osc}. The proposed mediators can be (i) a charged Higgs, (ii) colorless vector bosons ($W^\prime$), and (iii) scalar or vector leptoquarks (LQ). Scenarios with charged Higgs boson cannot accommodate $R_{D^{(\ast)}}$ due to the $B_c$-lifetime constraint discussed above, cf.~Fig.~\ref{fig:RD-RDst-1TeV}. On the other hand, $W^\prime$ models have difficulties to evade nontrivial LHC constraints on di-tau production~\cite{Greljo:2015mma,Faroughy:2016osc,Buttazzo:2017ixm}.~\footnote{These constraints can be avoided by considering a $W^\prime$ with couplings to right-handed neutrinos~\cite{Greljo:2018ogz}. However, in that case the interference term with the SM rate is not present, requiring rather large NP couplings. See also Ref.~\cite{Becirevic:2016yqi} for a similar proposal with scalar LQs.} The only minimalistic scenarios capable of explaining $R_{D^{(\ast)}}^{\mathrm{exp}}$ while satisfying other existing constraints are LQs. 

In Table~\ref{tab:states} we list the LQ states that can contribute to $b\to c\tau \bar{\nu}$ in terms of their SM quantum numbers, $(SU(3)_c,SU(2)_L,Y)$, with $Q=Y+T_3$~\cite{Dorsner:2016wpm}. From this Table we learn that very few possibilities are still viable:
%%%%%%%%%%%%%%%
\begin{itemize}
\item[•] The scalar LQ $S_1=(\mathbf{\bar{3}},\mathbf{1},1/3)$ generates, at the matching scale, the effective coefficients $C_{\ell q}^{(3)}$ and $C_{\ell e q u}^{(1)} =  -4 \, C_{\ell e q u}^{(3)}$, which induce nonzero values of $g_{V_L}$ and $g_{S_L} \approx -8.1 \, g_T$ at $\mu=m_b$, after accounting for RGE effects. Both of these Wilson coefficients can accommodate for $R_{D^{(\ast)}}$, as discussed in Sec.~\ref{sec:eft}. See also Ref.~\cite{Sakaki:2013bfa,Bauer:2015knc,Becirevic:2018uab,Marzocca:2018wcf}. 
\item[•] The scalar LQ  $R_2=(\mathbf{3},\mathbf{2},7/6)$ produces $C_{\ell e q u}^{(1)} =  4 \, C_{\ell e q u}^{(3)}$, which implies $g_{S_L} \approx 8.1 \, g_T$ at $\mu=m_b$. Note that this particularl correlation of scalar and tensor coefficients lies outside of the allowed regions in Fig.~\ref{fig:RD-RDst-1TeV}, where only real coefficients were considered. However, this problem can be overcome by considering complex Wilson coefficients~\cite{Sakaki:2013bfa,Hiller:2016kry,Becirevic:2018uab,Becirevic:2018afm}. 
\item[•] The vector LQ $U_1=(\mathbf{3},\mathbf{1},2/3)$ induces $C_{\ell q}^{(3)}$, which implies nonzero values of $g_{V_L}$ and $g_{S_R}$~\cite{Buttazzo:2017ixm}.
\end{itemize}
%%%%%%%%%%%%%%%

\noindent Note that some of the models which predict $g_{V_L}^\ell$ in Table~\ref{tab:states} cannot accommodate $R_{D^{(\ast)}}$, as shown in Table~\ref{tab:states}. This is the case because $g_{V_L}^\tau$ is constrained to be negative in these scenarios, providing an interference term with the wrong sign for $R_{D^{(\ast)}}$, cf.~Eq.~\eqref{eq:gvl-fit}~\cite{lqpaper}. Furthermore, a peculiarity of the viable models listed above is a different pattern of effective operators which imply very distinctive phenomenological consequences, as we will discuss in Section~\ref{sec:pheno}. 

Finally, a natural question is if the hint of New Physics in $R_{K^{(\ast)}}^{\mathrm{exp}}<R_{K^{(\ast)}}^{\mathrm{SM}}$ and $R_{D^{(\ast)}}^{\mathrm{exp}}>R_{D^{(\ast)}}^{\mathrm{SM}}$ can be simultaneously explained by the same NP scenario. Building such a model turns out be a very challenging task. The only mediator that can account for both anomalies at tree-level is the vector LQ $U_1$~\cite{Buttazzo:2017ixm}. The problem in this case is that a resulting model is nonrenormalible and an explicit UV-completion needs to be specified. This issue has been addressed in a series of papers with gauged models~\cite{Assad:2017iib,DiLuzio:2017vat,Bordone:2017bld,Barbieri:2017tuq,Blanke:2018sro}. A far more minimalistic approach, in terms of number of parameters, is to consider models containing two scalar LQs. These particles could arise, for instance, from $SU(5)$ GUT~\cite{Becirevic:2018afm} or composite Higgs models~\cite{Marzocca:2018wcf}.

\

%%%%%%%%%%%%%%%%%%%%%%
\begin{table}[htbp!]
\renewcommand{\arraystretch}{1.45}
\centering
\begin{tabular}{|c|c|c|cccc||c|}\hline
Field  & Spin & Quantum Numbers &   $g_{V_L}$ & $g_{S_R}$ & $g_{S_L}$  & $g_T$ & $R_{D^{(\ast)}}$\\ \hline\hline
$R_2$  & $0$ & $(\mathbf{3},\mathbf{2},7/6)$ & -- & -- & $\checkmark$   & $\checkmark$ & $\color{blue}\checkmark$ \\ 
$S_1$  & $0$ & $(\mathbf{\bar{3}},\mathbf{1},1/3)$ & $\checkmark$  & -- & $\checkmark$ &  $\checkmark$ & $\color{blue}\checkmark$\\ 
$S_3$  & $0$ & $(\mathbf{3},\mathbf{2},7/6)$ &  $\checkmark$ & -- & --  &  -- & {\color{red}\xmark}\\ \hline
$V_2$  & $1$ & $(\mathbf{\bar{3}},\mathbf{2},5/6)$ &  -- & $\checkmark$ & --  & -- & {\color{red}\xmark}\\ 
$U_1$ & $1$ & $(\mathbf{3},\mathbf{1},2/3)$ & $\checkmark$ & $\checkmark$ & -- & -- & $\color{blue}\checkmark$\\ 
$U_3$  & $1$ & $(\mathbf{3},\mathbf{3},2/3)$ &  $\checkmark$ & -- & --  & -- & {\color{red}\xmark}\\ \hline
\end{tabular}
\caption{ \sl \small List of LQ states contributing to the transition $b\to c\ell \bar{\nu}$ in terms of the SM quantum numbers, $(SU(3)_c,SU(2)_L,Y)$, and corresponding Wilson coefficients generated in Eq.~\eqref{leff:bc}. The last column summarizes the models which can accommodate $R_{D^{(\ast)}}$ while complying with other existing constraints~\cite{lqpaper}. See text for details.}
\label{tab:states} 
\end{table}
%%%%%%%%%%%%%%%%%%%%

\subsection{Phenomenological implications}
\label{sec:pheno}

We conclude this Section by discussing the phenomenological implications of $R_{D^{(\ast)}}$ and the strategies to disentangle the viable NP explanations, if the anomalies persist. A first possibility is to study LFUV ratios of other decay modes based on the transition $b\to c \ell\bar{\nu}$, such as $B_s\to D_s^{(\ast)} \tau \bar{\nu}$ and $B_c\to J/\psi \tau \bar{\nu}$. In particular, LHCb performed a first measurement of the ratio~\cite{Aaij:2017tyk}
%%%%%%%%%%%%%%%%
\begin{equation}
R_{J/\psi}=\dfrac{\mathcal{B}(B_c\to J/\psi \tau \bar{\nu})}{\mathcal{B}(B_c\to J/\psi \mu \bar{\nu})}=0.71 \pm 0.17 \pm 0.18\,,
\end{equation}
%%%%%%%%%%%%%%%%
which turns out to be again $\approx 2\sigma$ larger than the SM estimate, $R_{J/\psi}^{\mathrm{SM}}=0.23(1)$, obtained by combining LQCD and QCD sum rules results~\cite{RJpsipaper}. These independent measurements can be useful not only to corroborate/refute the excess found on $R_{D^{(\ast)}}$, but also to provide a complementary information about the underlying NP structure. For instance, as discussed in Sec.~\ref{sec:eft}, contributions in the form of (current)$\times$(current) operators predict the following equality,
%%%%%%%%%%%%%%
\begin{align*}
\dfrac{R_D}{R_D^{\mathrm{SM}}}=\dfrac{R_{D^\ast}}{R_{D^\ast}^{\mathrm{SM}}}=\dfrac{R_{J/\psi}}{R_{J/\psi}^{\mathrm{SM}}} = \dots\,,
\end{align*}
%%%%%%%%%%%%% 
\noindent  for all decay modes based on the same transition. On the other hand, scalar and/or tensor operators induce a distinct correlation between $P\to P^{\prime}$ and $P\to V$ decays, where $P^{(\prime)}$ and $V$ denote generic pseudoscalar and vector mesons, respectively. Another possibility proposed in the literature is to consider the angular observables of the decay $B\to D^\ast(\to D \pi) \tau \bar{\nu}$, which have different sensitivities to the operators in Eq.~\eqref{leff:bc}~\cite{Becirevic:2016hea,Ivanov:2017mrj}. Note, in particular, that a first measurement of the $\tau$-lepton polarization asymmetry in $B\to D^\ast \tau \bar{\nu}$ has been made by Belle, with large error bars~\cite{Hirose:2016wfn}. This observable is particularly sensitive to pseudoscalar and tensor contributions, but it is unaffected by interaction of the type $(V-A) \times (V-A)$~\cite{Azatov:2018knx}. 

These semileptonic observables described above offer an alternative test to the enhancement in $R_{D^{(\ast)}}$, without any assumption regarding the underlying NP structure. Model independent tests can also be obtained by using purely leptonic observables. As already anticipated in Sec.~\ref{sec:eft}, the interactions responsibles for the anomalies in $R_{D^{(\ast)}}$ must respect $SU(2)_L \times U(1)_Y$ gauge invariance. Writing the operators as in Eq.~\eqref{eq:su2-operators} at a scale $\mu \gg \mathcal{O}(v_{\mathrm{EW}})$ opens up the possibility to study the impact of electroweak corrections within an effective field theory context. It has been shown in Ref.~\cite{Feruglio:2016gvd,Feruglio:2017rjo,Cornella:2018tfd} that the operators generating the (current)$\times$(current) contributions to $R_{D^{(\ast)}}$ also induce sizable corrections to the $Z$-pole observables and to LFU tests in $\tau$-decays. On the other hand, the operators generating (pseudo)scalar and tensor contributions have the peculiarity of inducing chirality-enhanced contributions to $\mathcal{B}(h\to \tau\tau)$ and to the $\tau$-lepton anomalous magnetic moment, respectively, without inducing any sizable modification to the $Z$-pole observables and LFU tests in $\tau$-decays~\cite{Feruglio:2018fxo}. The ongoing experimental effort at Belle-II and LHC will offer the opportunity to check for an additional enhancement in these observables.   Finally, model-dependent predictions can also be obtained once a specific scenario is chosen. These include the signals at the LHC~\cite{Faroughy:2016osc} and many other flavor observables, such as lepton flavor violating decays, cf.~e.g.~Ref.~\cite{Glashow:2014iga,Becirevic:2016oho,Bordone:2018nbg}. I refer the reader to other talks presented at this conference in which these aspects have been covered in greater detail~\cite{othertalks}.

\section{Summary and perspectives}

In this overview we discussed the recent theoretical progress on the study of semileptonic $B$-meson decays. The focus of our discussion was (i) the dependence of the extraction of $|V_{cb}|$ from $B\to D^\ast l \bar{\nu}$ decays on the parameterization of the underlying form factors, and (ii) the possible interpretations of the very intriguing hints of LFU in charged currents:

\begin{itemize}
\item[•] While the recent reanalysis of $B\to D^\ast l \bar{\nu}$ data from Belle pointed out underestimated systematic uncertainties in the extraction of $|V_{cb}|^{\mathrm{excl.}}$~\cite{Bigi:2017njr,Grinstein:2017nlq}, the $V_{cb}$ puzzle still remains since both form-factor parameterizations provide an equally good description of current data. The solution to this puzzle will come in the future with Belle-II and with LQCD data on the slope of the form-factors near zero recoil. 
\item[•] After summarizing the status of SM predictions, we discussed the implications of the LFU violation observed in $R_{D^{(\ast)}}$. By using an effective field theory description we argued that the operator of type $(V-A)\times (V-A)$ and the tensor one can accommodate current discrepancies, while the purely scalar ones cannot. These scenarios can be experimentally tested (i) by measuring similar LFU ratios based on the same quark transition, (ii) by studying the angular asymmetries of $B\to D^{(\ast)} \tau \bar{\nu}$ decays, and (iii) by studying purely leptonic observables, which are related to the LFU breaking effects via RGE effects. We have also argued that LQ bosons are the best candidates to explain these anomalies and we summarized the viable proposals made in the literature.
\end{itemize} 

\section*{Acknowledgments}

I would like to thank the organizers for the invitation and especially to Damir Be\v cirevi\' c, Svjetlana Fajfer, Ferruccio Feruglio, Nejc Ko\v snik and Paride Paradisi from whom I learned so much about the subject of this talk. This project has received support by the European Union's Horizon 2020 research and innovation programme under the Marie Sklodowska-Curie grant agreement N$^\circ$~674896.

\end{document}